\documentclass[prb,twocolumn]{revtex4}
\usepackage{graphicx}

\begin{document}

\title{Calculation of conduction-to-conduction and valence-to-valence 
transitions between bound states in (In,Ga)As/GaAs quantum dots} 
\author{Gustavo A. Narvaez} 
\altaffiliation{Current address: Eclipse Energy Systems, Inc., St. Petersburg,
Florida 33710}
\email{gnarvaez@eclipsethinfilms.com}
\author{Alex Zunger}
\email{alex_zunger@nrel.gov}
\affiliation{National Renewable Energy Laboratory, Golden, Colorado 80401}
\date{\today}

\begin{abstract}
  We have calculated the conduction-to-conduction and valence-to-valence
  absorption spectrum of bound states in (In,Ga)As/GaAs quantum dots charged
  with up to three electrons or holes. Several features emerge:
  (i) In pure (non-alloyed) InAs/GaAs dots, the $1S$-$1P_1$ and $1S$-$1P_2$
  conduction intraband transitions are fully in-plane polarized along $[1\bar
  10]$ and $[110]$, respectively, while valence transitions are weakly
  polarized because the hole {\em P} states do not show any in-plane
  preferential orientation.
  (ii) In {\em alloyed} In$_{0.6}$Ga$_{0.4}$As/GaAs dots the $[110]$ and
  $[1\bar 10]$ polarization of the corresponding $1S$-$1P$ conduction
  intraband transitions is weakened since the two $1P$ states are mixed by
  alloy fluctuations.  The polarization of valence intraband transitions is
  insensitive to changes in alloy fluctuations.
 (iii) For light polarized along [001], we find a strong valence-to-valence
 transition that involves a weakly confined hole state with predominant
 light-hole character. 
  (iv) When charging the dots with a few electrons, the conduction intraband
  transitions display spectroscopic shifts of $\sim 1$-$2\;{\rm meV}$. These
  shifts are a result of correlation effects (captured by
  configuration-interaction) and not well described within the Hartree-Fock
  approximation.
  (v) When charging the dots with holes, valence intraband spectra are more
  complex than the conduction intraband spectra as hole states are strongly
  affected by spin-orbit coupling, and configuration mixing is more
  pronounced.  Spectroscopic shifts can no longer be identified unambiguously.
  These predictions could be tested in {\em single-dot} spectroscopy of {\em
    n}-doped and {\em p}-doped quantum dots.
\end{abstract}

\maketitle

\section{Brief background and scope}

Doping quantum dots {\em n}-type ({\em p}-type) followed by infrared-light
excitation leads to conduction-to-conduction (valence-to-valence) intraband
excitations between confined states. 
In early experiments, intraband transitions were studied for {\em quantum
  wells},\cite{levine_JAP_1993} where in the case of {\em n}-type doping the
selection rules lead to allowed transitions only for light polarized normal to
the well,\cite{west_APL_1985,levine_APL_1986,liu_APL_1998} whereas {\em
  p}-type samples afforded transitions allowed in normal incidence. Instead,
{\em n}-doped {\em quantum dots} offered the possibility of allowed intraband
transitions for light incident normal to the plane of the dots.  Thus, quantum
dots became the focus of much experimental work devoted to study the intraband
optical
transitions.\cite{drexler_PRL_1994,fricke_EPL_1996,sauvage_PRB_1998,sauvage_JAP_1998,chu_APL_1999,chu_APL_2000,sauvage_APL_2001,goede_PRB_2001,hameau_intra,aslan_APL_2003,zibik_JAP_2006}
For example, measurements of optical intraband transitions in (In,Ga)As/GaAs
quantum dots charged with electron and holes have been recently performed by
Zibik {\em et al.}\cite{zibik_PhysicaE_2005}, who measured
conduction-to-conduction transitions, and Preisler {\em et
  al.}\cite{preisler_PRB_2005} who have measured valence-to-valence intraband
absorption. 

On the other hand, calculations of the intraband optical
properties are often based on model\cite{wojs_PRB_1996,leburton_JJAP_1999,zhang_APL_2004} or
${\bf k}\cdot{\bf p}$\cite{jiang_APL_1997} approaches and not high-level
atomistic approaches. Here, we address this issue by calculating
conduction-to-conduction and valence-to-valence intraband optical absorption
of (In,Ga)As/GaAs dots charged with up to three electrons or holes.
We do not intend to survey the effects of size, shape, and composition on the
intraband transitions of the dots. We focus primarily on (i) the
effects of alloy fluctuations on the polarization of the intraband transitions,
and (ii) the differences between the absorption spectra of {\em n}-doped and
{\em p}-doped dots as well as the spectroscopic shifts induced by
charging. 
We use a combined approach to calculate the intraband absorption spectra in
which we find the electron and hole single-particle states of the dots with an
atomistic pseudopotential method and solve the many-particle states of charged
dots within a configuration-interaction approach. The advantage of this
approach over simplified methods is that it naturally includes (i) the correct
symmetry of the dot; (ii) strain and alloying effects; (iii) multiband and
multivalley coupling; and (iv) spin-orbit interaction.
Our approach is purely electronic and does not include polaron
(electron-phonon coupling) effects.
We illustrate our findings with a prototypical lens-shaped
In$_{0.6}$Ga$_{0.4}$As/GaAs dot with diameter $b=252\;${\AA} and height
$h=35\;${\AA}. As a benchmark, for dots charged with a single carrier, we
provide results for a pure non-alloyed InAs/GaAs dot with the same size. 
While so far intraband experiments have focused on ensemble of dots with
different degrees of homogeneity, which broadens the observed transitions, we
offer predictions that could be probed in single-dot spectroscopy of {\em n}-
or {\em p}-doped dots. In addition, as the control of doping carriers is
difficult, i.e in an ensemble of doped dots some have a single electron while
other have two or none, we discuss the intraband spectra upon increasing the
carriers one at a time.

%%%%%%%%%%%%%%
%   Method   %
%%%%%%%%%%%%%%
%
%
\section{Method: Pseupotential approach and configuration-interaction}
\label{Sec_method}

We calculate the single-particle electron and hole energy levels of the
self-assembled dot within an atomistic, pseudopotential-based
method:\cite{zunger_pssb_2001} The wave function $\psi_j$ and energy ${\cal
  E}_j$ are solutions of the atomistic single-particle
Schr\"odinger equation

\begin{equation}
\label{SP.Shrodinger}
\Big[-\frac{1}{2}\nabla^2+V_{SO}+\sum_{l,\alpha}\,v_{\alpha}({\bf R}-{\bf R}_{l,\alpha})\Big]\psi_j={\cal
  E}_j\,\psi_j 
\end{equation}

\noindent where $v_{\alpha}$ is the screened
pseudopotential for atom of type $\alpha$ (In,Ga,As) with position ${\bf
  R}_{l,\alpha}$ within the dot or barrier, and $V_{SO}$ is a non-local
pseudopotential that accounts for the spin-orbit
interaction.\cite{williamson_PRB_2000} These pseudopotentials are carefully
fitted to bulk GaAs, InAs, and (In,Ga)As alloys, thus removing the
local-density-approximation (LDA) errors. The basis in which we expand
$\psi_j$ to solve Eq. (\ref{SP.Shrodinger}) is a linear combination of {\em
  full-zone} Bloch bands of the underlying solids.\cite{wang_PRB_1999} Thus,
this method incorporates multi-band and multivalley coupling, band
non-parabolicity, and spin-orbit effects, as well as the effects of the
underlying strain in the dot and barrier.

To solve for the many-particle states $\{\Psi_{\nu}({\cal N}),\,E_{\nu}({\cal
  N})\}$ of the dot with ${\cal N}$ carriers, where ${\cal N}=N_e$ electrons
or $N_h$ holes, we use a configuration-interaction approach with screened
direct ($J$) electron-electron and hole-hole Coulomb interaction and exchange
($K$).\cite{franceschetti_PRB_1999} This method has been recently applied to
the calculation of electronic and optical properties of (In,Ga)As/GaAs dots
such as electron and hole charging\cite{he_charging}, radiative lifetimes of
neutral and charged excitons,\cite{narvaez_PRB_2005} relaxation times of
electrons due to electron-hole Auger scattering,\cite{narvaez_PRB_2006b} and
fine-structure splittings of neutral and charged
excitons.\cite{bester_PRB_2003}

At low temperatures such that only the ground state $\Psi_0({\cal N})$ of
the ${\cal N}$-carrier dot is significantly occupied, the optical absorption
spectrum for light polarized along $\hat{\bf e}$ is given by

\begin{equation}
\label{Eq_absorption}
I(\hbar\omega;\hat{\bf e})=\sum_{{\nu}^{\prime}}|\langle\Psi_{\nu^{\prime}}|\hat{\bf e}\cdot{\bf p}|\Psi_{0}\rangle|^2\delta(E_{\nu^{\prime}}-E_{0}-\hbar\omega).
\end{equation}

\noindent In the results we present subsequently, we have phenomenologically 
broadened the spectra with a Gaussian of width $\sigma=0.25\;{\rm meV}$. Such
a broadening has also been used in other simulations of optical absorption in
the literature.

For dots with {\em cylindrical} symmetry, in which the projection of the
angular momentum along the cylindrical axis ($\hat{\bf z}$) is a good quantum
number, the selection rules for intraband transitions are the following. For
$\hat{\bf e}\parallel [100]$ or $[010]$ transitions are allowed between states 
such that $\Delta L_z=\pm 1$, while for $\hat{\bf e}\parallel [001]$ only
transitions with $\Delta L_z=0$ are allowed.

We consider in our calculations prototypical lens-shaped
In$_{0.6}$Ga$_{0.4}$As/GaAs dots with base diameter $b=252\;${\AA} and height
$h=35\;${\AA}. 
As a benchmark, in the case of dots charged with a single carrier we also
consider a pure non-alloyed InAs/GaAs dot of the same size.\cite{note-900}
Note that the detailed experimental characterization of the shape, size, and
composition of the alloyed (In,Ga)As/GaAs dots probed optically is scarce. The
prototypical alloyed dot we consider here presents properties like excitonic
gap; electron and hole single-particle energy spacings; and binding energies
of neutral and excited excitons in excellent agreement with available
data.\cite{narvaez_PRB_2005}

%%%%%%%%%%%%%%%%%%%%%%%%%%%%%%%%%%%%%%%
%                                     %
%   Single electron and single hole   %
%                                     %
%%%%%%%%%%%%%%%%%%%%%%%%%%%%%%%%%%%%%%%
%
%
\section{Results for single-particle electron and hole levels}

\begin{figure}
\includegraphics[width=8.5cm]{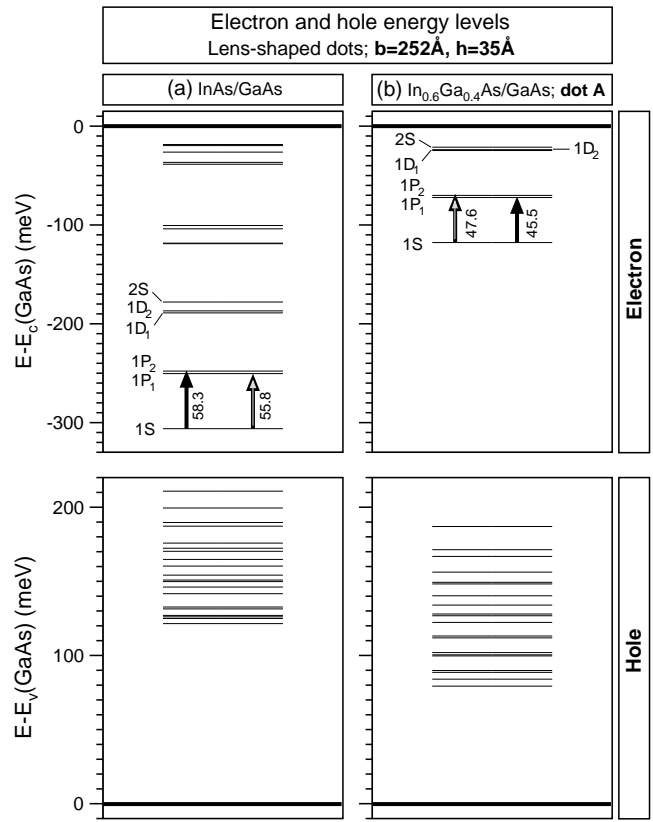}
\caption{{\label{Fig_m1}}Electron (top) and hole (bottom) energy levels in
  lens-shaped (In,Ga)As/GaAs dots with base diameter $b=252\;${\AA} and height
  $h=35\;${\AA} relative to the GaAs barrier conduction band minimum $E_c{\rm
    (GaAs)}=-4.093\;{\rm eV}$ (w.r.t vacuum) and its valence band maximum
  $E_v{\rm (GaAs)}=-5.620\;{\rm eV}$ (w.r.t vacuum). The energy levels are
  more confined in the pure (non-alloyed) InAs/GaAs dot (a) than in the
  alloyed In$_{0.6}$Ga$_{0.4}$As/GaAs dot (b).  For holes, we show the first
  40 confined states. Black and grey arrows indicate, respectively, the lowest
  intraband conduction transition for polarization along $[110]$ and $[1\bar 1
  0]$; while numbers show the transition energy in {\rm meV}.}
\end{figure}

\subsection{Electron levels}

Bound states can be labeled by their leading orbital character and
approximately arranged into {\em shells}:
$\{1S;\,1P_1,1P_2;\,1D_1,1D_2,2S;\,\ldots\}$.  For the dot size considered in
our calculation, a pure (non-alloyed) dot confines 15 states arranged in five
shells [Fig. \ref{Fig_m1}(a)], two more shells than in its alloyed counterpart
which binds 10 states [Fig. \ref{Fig_m1}(b)]. The larger number of confined
levels in a pure dot is due to a larger strain-modified conduction band offset
than that in an alloyed dot.
% S-shell properties
%
In the pure InAs/GaAs dot [Fig. \ref{Fig_m1}(a)] the $1S$ state is located
$306\;{\rm meV}$ below the conduction band minimum (CBM) of the GaAs barrier,
while in the alloyed dots [Fig.  \ref{Fig_m1}(b)] is $\sim 118\;{\rm meV}$
below the GaAs CBM. (This value changes by a few {\rm meV}s depending on the
random alloy fluctuations in the dot.) These energies of the $1S$ state
relative to the GaAs CBM set the cutoff for conduction-to-conduction 
intraband transitions between bound states.
% P-shell properties
%
The $P$-shell consists of two non-degenerate states $1P_1$ and $1P_2$. The
origin of this splitting is atomistic\cite{bester_PRB_2005} and a consequence
of the underlying $C_{2v}$ symmetry of the (pure, nonalloyed) dots, which is
lower than $C_{\infty v}$ symmetry of the macroscopic (lens) shape, which is
normally assumed in continuum effective-mass models. In our calculation we
consider perfectly cylindrical dots; yet, the $P$-$P$ splitting is as large as
$2\;{\rm meV}$ in pure dots. Piezoelectricity\cite{bester_newRC} and
non-cylindrical shape\cite{wang_PRB_1999} further contribute to this $P$-$P$
splitting. In addition, each of these $P$ states present a nearly equal
mixture of $L_z=\pm 1$ components contrary to the axially-symmetric case in
which each state has a well defined $L_z$ component.
% Alloy fluctuations in P-shell
%
In the pure (non-alloyed) dot $1P_1$ is oriented along $[1\bar 10]$ and $1P_2$
is oriented along $[110]$ [Fig. \ref{Fig_0}(a)]. In alloyed dots the
symmetry of the dots is lower than $C_{2v}$ due to random alloy fluctuations.
In this case, not only $[110]$ and $[1\bar 10]$ are mixed and different
disorder realizations (fluctuations) change the magnitude of the P-P splitting
by $1$-$3\;{\rm meV}$ but, more remarkably, alloy fluctuations affect the
in-plane orientation (polarization) of the $P$ states, as is shown in Fig.
\ref{Fig_0}. Or, equivalently, alloy fluctuations change the relative phase
$\phi_{\pm}$ of the $L_z=\pm 1$ components in the $1P_1$ and $1P_2$ states,
which results in different in-plane orientations (polarizations) of these
states. For instance, in dot A we have $\phi_+\simeq 0$ and $\phi_-\simeq
\pi/2$, while for dot C we have $\phi_{+}\simeq\phi_{-}\simeq 0$.
% D-shell properties
% 
In turn, the $D$-shell consists of non-degenerate $1D_1$, $1D_2$, and $2S$
states. States $1D_1$ and $1D_2$ show a nearly even mixture of $L_z=\pm 2$
components. Depending on alloy fluctuations, state $2S$ can also have sizeable
$L_z=\pm 2$ components, thus making it not possible in those cases to assign a
leading orbital character to these $D$-shell states.

\subsection{Hole levels}

\begin{figure}
\includegraphics[width=8.5cm]{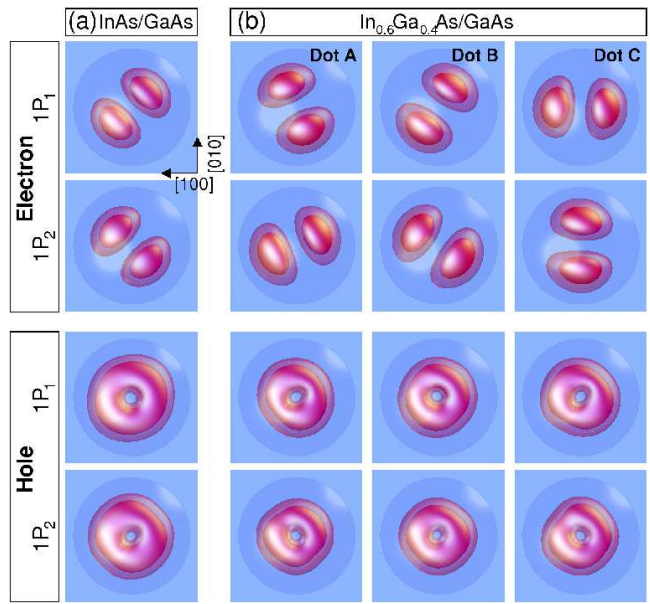}
\caption{{\label{Fig_0}}(Color online) Isosurface representation of the
  electron (top) and hole (bottom) wavefunctions of states $1P_1$ and $1P_2$
  in a pure non-alloyed InAs/GaAs dot (a), and three
  In$_{0.6}$Ga$_{0.4}$As/GaAs dots (b) with different alloy fluctuations. Dots
  are lens-shaped and have the same size: $b=252\;${\AA} and
  $h=35\;${\AA}.}
\end{figure}

Both non-alloyed and alloyed dots confine a large ($M_h > 20$) number of
single-particle levels. Due to the multi-band nature of these hole states and
for flat dots like the one we consider here, only low-lying states present
shell structure that is less pronounced, i.e. larger $P$-$P$ and $D$-$D$
splittings, than in the electron case. For these states, one can still use
their leading $S$, $P$, $D$ orbital character to identify them. Higher lying
states show heavy mixing of orbital character.
% 1S states
%
For pure and alloyed dots the $1S$ hole state is located, respectively,
$211\;{\rm meV}$ and $186\;{\rm meV}$ above the valence band maximum of the
GaAs barrier; see Fig. \ref{Fig_m1}.  These values are the cutoff for
valence-to-valence intraband transitions between bound states. In addition,
the $P$ states are not oriented along a preferential in-plane direction [Fig.
\ref{Fig_0}].

\section{Intra-conduction and intra-valence transitions in dots with a single carrier}

Assuming that only the $1S$ state is occupied by doping, for in-plane
polarized light we expect intra-conduction transitions between bound states
that satisfy $\Delta L_z=\pm 1$ when the dot is occupied by a single electron.
The lowest-energy transitions correspond to $1S$-$1P_1$ and $1S$-$1P_2$
(indicated by arrows in Fig.  \ref{Fig_m1}).  The orientation (polarization)
of the $1P$ states determines the polarization properties of these
transitions:

(i) For pure (non-alloyed) InAs/GaAs dots, we expect that for polarization
$\hat{\bf e}\parallel [1\bar 10]$ only transition $1S$-$1P_1$ to be allowed,
while for $\hat{\bf e}\parallel [110]$ only $1S$-$1P_2$.

(ii) For alloyed dots, alloy fluctuations dictates the orientation of states
$1P_1$ and $1P_2$ [Fig. \ref{Fig_0}(b)]. Thus, we expect a strong dot-to-dot
dependence of the polarization properties of the two conduction intraband
transitions. In particular, for the in-plane polarization $\hat{\bf
  e}\parallel [1\bar 10]$ we expect both transitions $1S$-$1P_1$ and
$1S$-$1P_2$ to be allowed, with intensities

\begin{eqnarray}
|\langle 1S|\hat{\bf e}\cdot{\bf p}|1P_1\rangle|^2 \propto 1-\sin(\phi_{+}-\phi_{-}) \\
|\langle 1S|\hat{\bf e}\cdot{\bf p}|1P_2\rangle|^2 \propto
|1+\sin(\phi_{+}-\phi_{-})| .
\end{eqnarray}

\begin{figure}
\includegraphics[width=8.5cm]{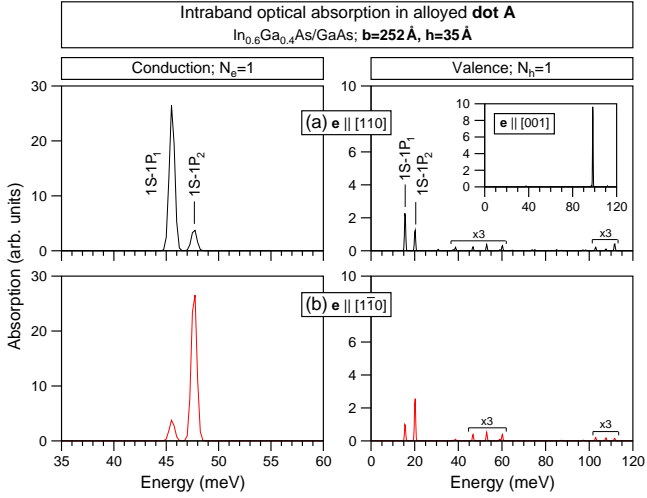}
\caption{{\label{Fig_1}}(Color online) Intraband absorption spectrum for in-plane
  polarization $\hat{\bf e}\parallel [110]$ (a) and $\hat{\bf e}\parallel
  [1\bar 10]$ (b) in an In$_{0.6}$Ga$_{0.4}$As/GaAs dot (dot A) charged with a
  single electron (left panels) and a single hole (right panels). For these
  in-plane polarizations, conduction-to-conduction transitions are nearly an order
  of magnitude more intense that valence-to-valence transitions (note different ordinate
  scales). Inset: Valence intraband transition for light polarized out-of-plane
  $\hat{\bf e}\parallel [001]$. The strong feature corresponds to the
  $1S_{hh}$-$1S_{lh}$ transition.
}
\end{figure}

%
% In-plane transitions
%
\subsection{In-plane transitions} 

In a pure (non-alloyed) InAs/GaAs charged with one electron the lowest
transition $1S$-$1P_1$ appears at $56\;{\rm meV}$ and is {\em fully} polarized
along $[1\bar 10]$, while the second transition $1S$-$1P_2$ is located at
$58\;{\rm meV}$ and is {\em fully} polarized along $[110]$. In contrast, in
alloyed dots [Figs. \ref{Fig_1}(a) and \ref{Fig_1}(b)] we only find partial
in-plane polarization: For dot A we find a transition at $46$ and another at
$48\;{\rm meV}$ corresponding; the lowest-energy transition $1S$-$1P_1$ is
partially ($\sim 75\,\%$) polarized along $\hat{\bf e}=[1 1 0]$ while the
higher-energy $1S$-$1P_2$ is partially ($\sim 75\%$) polarized along $\hat{\bf
  e}=[1\bar 1 0]$.

The valence intraband transitions have a smaller oscillator strength than
those of the conduction transitions. For holes, as in the case of electrons,
we find two strong transitions corresponding to the $1S_h$-$1P_h$ transitions.
However, these transitions are weakly in-plane polarized because the hole $P$
states do not show any preferential in-plane orientation. In addition, we also
find weaker transitions in the interval $40$-$60\;{\rm meV}$, which arise from
the mixing between the $D$-shell and $P$-shell hole states. For dot A, the
lowest valence intraband transition shows a higher intensity when light is
polarized along the $[1\bar1 0]$ direction, as in the experiment of Preisler
{\em et al.}\cite{preisler_PRB_2005}

\begin{figure}
\includegraphics[width=8.5cm]{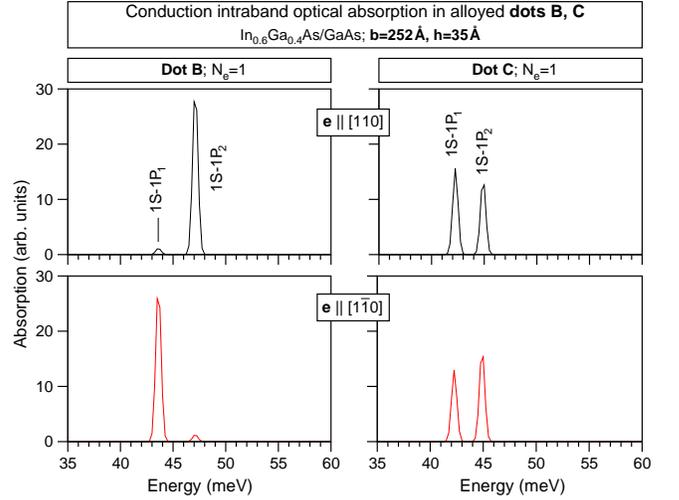}
\caption{{\label{Fig_2}} {\em Idem} Fig. \ref{Fig_1}(a) and
  Fig. \ref{Fig_1}(b) for two In$_{0.6}$Ga$_{0.4}$As/GaAs dots with the same
  size as dot A but with different random alloy fluctuations.}
\end{figure}

% Alloy fluctuation effects
% 
Calculations based on simplified models are not capable of explicitly
introducing alloy fluctuations.  Instead, these fluctuations are naturally
included within our atomistic approach.
Figure \ref{Fig_2} illustrates the effect of random alloy fluctuations on
the polarization properties of the conduction intraband transitions in
In$_{0.6}$Ga$_{0.4}$As/GaAs dots with the same size as dot A but different
random alloy fluctuations: (i) Under $[1\bar 10]$ polarization, dot B presents
transition $1S$-$1P_1$ nearly fully polarized and transition $1S$-$1P_2$
nearly forbidden; conversely, for $\hat{\bf e}\parallel [1 10]$ transition
$1S$-$1P_2$ is nearly fully polarized and $1S$-$1P_1$ nearly forbidden. These
polarization properties are {\em switched} when compared to dot A. In
addition, that the lowest conduction intraband transition in {\em dot B} is
mainly polarized along $[1\bar 10]$ in both is in agreement with the
experiment of Zibik {\em et al.}\cite{zibik_PhysicaE_2005}
(ii) In contrast, dot C presents both transitions allowed for polarizations
$[110]$ and $[1\bar 10]$, with very small in-plane polarization anisotropies.
We find that transition $1S$-$1P_1$ is polarized along [100], with transition
$1S$-$1P_2$ forbidden, and that for $\hat{\bf e}\parallel [010]$ transition
$1S$-$1P_2$ is allowed while $1S$-$1P_1$ is forbidden.

%
% Out-of-plane polarization
%
\subsection{Out-of-plane polarization---intra-valence transitions}

The inset of Fig. \ref{Fig_1}(a) shows the valence intraband transition for
$\hat{\bf e}\parallel [001]$. We find a strong feature originated from the
$1S_{hh}$-$1S_{lh}$ transition, which involves a weakly confined, highly
excited hole state with predominant light-hole character. This transition is
nearly three times as intense as the in-plane valence transitions.  This
transition is consistent with the selection rule $\Delta L_z=0$ for this light
polarization.

%%%%%%%%%%%%%%%%%%%%%%%%%%%%%%%%%%%%%%%%%%%%%
%                                           %
% Electron and hole occupation dependence   %
%                                           %
%%%%%%%%%%%%%%%%%%%%%%%%%%%%%%%%%%%%%%%%%%%%%
%
%
%
\section{Intra-conduction and intra-valence transitions in dots with a few carriers}

We now study the conduction and valence intraband transitions for ${\cal
  N}=2$, and $3$ carriers occupying the dot.

% Theory
%
The energy of conduction and valence band transitions in the presence of $N_e$
electron or $N_h$ holes is dictated by differences in total energies. [See Eq.
(\ref{Eq_absorption}).] Therefore, we expect spectroscopic shifts of the
transitions upon charging. 
% Spectroscopic shift
%
We define the intraband spectroscopic shift $\Delta_{if}$ of transition
$\Psi_i\rightarrow\Psi_f$ upon adding a carrier as

\begin{equation}
\Delta_{if}({\cal N})=\hbar\omega_{if}({\cal N})-\hbar\omega_{if}({\cal N}-1).
\end{equation}

\noindent The energy of a many-particle state $|\Psi_{\nu}\rangle$ can 
be expressed as superposition of a Hartree-Fock term [$E^{\,(HF)}$] and
correlation component ($\delta$); namely,

\begin{equation}
E^{(CI)}_{\nu}({\cal N}) = E^{\,(HF)}_{\nu}({\cal N}) + \delta_{\nu}({\cal N}).
\end{equation}

\noindent For the conduction-to-conduction intraband transition energies and shifts, 
we compare results for the conduction intraband transition energies and shifts
obtained within the Hartree-Fock approximation and CI calculations.

\begin{figure}
\includegraphics[width=8.5cm]{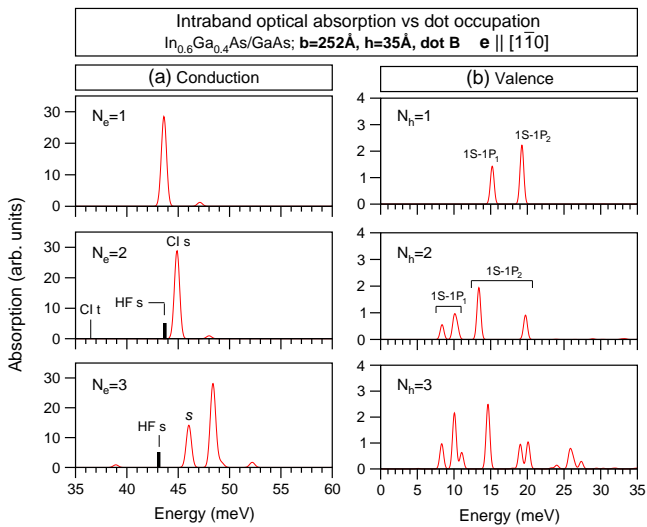}
\caption{{\label{Fig_3}}(Color online) For in-plane polarization
  $\hat{\bf e}\parallel [1\bar 10]$, intraband conduction-to-conduction (a)
  and valence-to-valence (b) transitions calculated at the CI level versus
  number of electrons $N_e$ and holes $N_h$ in the dot. Black, short bars in
  panel (a) for $N_e=2$ and $3$ show the $1S$-$1P$ transition energies
  predicted within the Hartree-Fock approximation---Eqs.  (\ref{HF_shift_N=2})
  and (\ref{HF_shift_N=3}), respectively. For $N_e=2$, the HF singlet (${\rm
    HF\;s}$) and the configuration-interaction singlet (${\rm CI\;s}$) and
  triplet (${\rm CI\;t}$) are indicated. The latter being forbidden.}
\end{figure}

%%%%%%%%%%%%%%%%%%%%%%%%
%   Intraband vs N_e   %
%%%%%%%%%%%%%%%%%%%%%%%%
%
%
\subsection{Intra-conduction transitions vs $N_e$}

We present Hartree-Fock (HF) expressions for the intraband transition energies
when the dot is charged with $N_e=1,\,2,\,3$ and then compare to
configuration-interaction calculations. To illustrate our findings, we
consider {\em dot B}, which has $1P_1$ and $1P_2$ states that are nearly fully
polarized along $[1\bar 10]$ and $[110]$, respectively, and focus on light
polarized along $[1\bar 10]$.

%
%   N_e=1
%
{\em $N_e=1$:} The ground state is $|\Psi_0\rangle=|e^1_0\rangle$ and under
IR light the final state is $|e^1_1\rangle$.  (Both states are twofold
degenerate.) For this occupation, there are no many-body effects and the energy of
the initial and final state are simply given by

\begin{eqnarray}
E_0(1) & = & {\cal E}^{\,(e)}_{1S} \\
E_f(1) & = & {\cal E}^{\,(e)}_{1P_1} .
\end{eqnarray}

\noindent The intraband transition energy is

\begin{equation}
\hbar\omega_{if}(1) = E_f-E_i = {\cal E}^{\,(e)}_{1P_1} - {\cal
  E}^{\,(e)}_{1S} .
\end{equation}

%
%   N_e=2
%
{\em $N_e=2$:} The closed-shell (nondegenerate) state
$|\Psi_0\rangle=|e^2_0\rangle$ is the ground state, and there are four
possible final states originating from $|e^1_0e^1_1\rangle$. At the
single-particle level the four final states are degenerate, but within the HF
approximation these states split in a triplet ($t$) and a singlet ($s$):
$|e^1_0e^1_1\rangle_{t}$ and $|e^1_0e^1_1\rangle_{s}$.  The splitting between
these states is due to the electron-electron exchange interaction, between an
electron in $|S\rangle$ and an electron in $|P_1\rangle$, and equals
$2\,K^{\,(ee)}_{SP_1}$. Under IR light excitation, only the singlet is an
allowed final state. Thus, we have the following energies for the initial and
final state:

\begin{eqnarray}
E_0(2) & = & 2\,{\cal E}^{\,(e)}_{1S} + J^{\,(ee)}_{SS} + \delta_{0}(2) \\
E_f(2) & = & {\cal E}^{\,(e)}_{1P_1}+{\cal E}^{\,(e)}_{1S} + J^{\,(ee)}_{SP_1} +
K^{\,(ee)}_{SP_1} + \delta_{f}(2).
\end{eqnarray}

\noindent This results in a $1S$-$1P$ transition energy in the presence of an
extra electron that can be written as 

\begin{equation}
\label{N=3_omega}
\hbar\omega_{if}(2) = {\cal E}^{\,(e)}_{1P_1} - {\cal E}^{\,(e)}_{1S} +
J^{\,(ee)}_{SP_1} + K^{\,(ee)}_{SP_1} - J^{\,(ee)}_{SS} + \big[\delta_f(2)-\delta_0(2)\big].
\end{equation}

\noindent  Here, $J^{(ee)}_{SS}$ and $J^{(ee)}_{SP_1}$ are, respectively, 
the direct Coulomb interaction between two electrons in $|S\rangle$, and one
electron in $|S\rangle$ and another in $|P_1\rangle$. For dot B our atomistic
calculation predicts $J^{\,(ee)}_{SS}=20.8\;{\rm meV}$ and
$J^{\,(ee)}_{SP_1}=16.6\;{\rm meV}$, and $K_{SP_1}=4.3\;{\rm meV}$.

Our CI calculations for $N_e=2$ are shown in Fig. \ref{Fig_3}(a). We find that
the four states arising from configuration $e^1_0e^1_1$ indeed split in a
singlet and triplet and, as expected, only the intraband transition from the
ground-state to the singlet is allowed---strong feature at $\sim 45\;{\rm
  meV}$. The Hartree-Fock predicted transition energy [first five terms in Eq.
(\ref{N=3_omega})] for dot B appears as a short black bar in Fig.
\ref{Fig_3}(a) and is $43.7\;{\rm meV}$, while CI predicts $44.9\;{\rm meV}$,
so the correlation correction [terms in brackets in Eq. (\ref{N=3_omega})] is
$1.2\;{\rm meV}$.

The spectroscopic shift between the transitions in $N_e=1$ and $2$ is thus
given by

\begin{eqnarray}
\label{HF_shift_N=2}
\Delta(2) & = & J^{\,(ee)}_{SP_1} - J^{\,(ee)}_{SS} + K^{\,(ee)}_{SP_1} + \big[\delta_f(2)-\delta_0(2)\big].
\end{eqnarray}

\noindent The Hartree-Fock component of this shift is $0.1\;{\rm
  meV}$, whereas our CI calculation reveals a blue-shift of
$\Delta(2)=1.3\;{\rm meV}$; i.e., $1.2\;{\rm meV}$ from correlation effects
[terms in bracket in Eq.  (\ref{HF_shift_N=2})]. This shows correlation
effects determine the spectroscopic shift.

%
%   N_e=3
%
{\em $N_e=3$:} The ground state of the initial state is $|e^2_0e^1_1\rangle$
(twofold degenerate) and in addition to the final states originating from the
$P$ shell: $|e^1_0e^1_1\rangle$ and $|e^1_0e^1_2\rangle$, one now has six
states arising from transitions to the $D$ shell: $|e^2_0e^1_3\rangle$,
$|e^2_0e^1_4\rangle$, and $|e^2_0e^1_5\rangle$.
The energy of the ground state and the $P$-shell derived final are the following:

\begin{eqnarray}
E_0(3) & = & 2\,{\cal E}^{\,(e)}_{1S} + {\cal E}^{\,(e)}_{1P_1} + J^{\,(ee)}_{SS}
+ 2\,J^{\,(ee)}_{SP}-K^{\,(ee)}_{SP} + \delta_{0}(3) \\
E_{fP}(3) & = & 2\,{\cal E}^{\,(e)}_{1P_1}+{\cal E}^{\,(e)}_{1S} +
2\,J^{\,(ee)}_{SP} + J^{\,(ee)}_{PP} +\delta_{f}(3) .
\end{eqnarray}

\noindent Here, $J^{\,(ee)}_{PP}$ is the direct Coulomb interaction of two
electrons occupying $|P_1\rangle$. For dot B, $J^{\,(ee)}_{PP}=16.0\;{\rm
  meV}$. 
The resulting $1S$-$1P_1$ intraband transition energy is

\begin{equation}
\hbar\omega_{0f}(3) = {\cal E}^{\,(e)}_{1P_1} - {\cal E}^{\,(e)}_{1S} +
J^{\,(ee)}_{P_1P_1} - J^{\,(ee)}_{SS}+K^{\,(ee)}_{SP_1}
+\big[\delta_{f}(3)-\delta_{0}(3)\big] .
\end{equation}

\noindent For dot B, the HF component amounts to $43.1\;{\rm meV}$ [black bar in Fig. 
\ref{Fig_3}(a)] and the term due to correlations (in brackets) amounts to
$2.9\;{\rm meV}$.

The spectroscopic shift between the $1S$-$1P_1$ transitions in $N_e=2$ and $3$
is

\begin{equation}
\label{HF_shift_N=3}
\Delta(3) = J^{\,(ee)}_{P_1P_1} - J^{\,(ee)}_{SP_1} + 
\big[\delta_{f}(3)-\delta_{f}(2)-\delta_{0}(3)+\delta_{0}(2)\big].
\end{equation}

\noindent The HF part is a red-shift of $\Delta(3)=-0.6\;{\rm meV}$. 
The value predicted by EMA is nearly twice as big; namely,
$[\Delta(3)]_{EMA}=-1.3\;{\rm meV}$.
% Actual results
%
Yet, our CI calculation shows [Fig. \ref{Fig_3}(a) for $N_e=3$] that the
transition is actually {\em blue-shifted} by $1.1\;{\rm meV}$ with respect to
that in $N_e=2$. The correlations contribution [in brackets in Eq.
(\ref{HF_shift_N=3})] being $1.7\;{\rm meV}$. Note also that the intensity of
transition $1S$-$1P_1$ is bleached (reduced) by nearly $50\,\%$ as a
consequence of Pauli blocking---one electron occupies $|1P_1\rangle$.

Regarding the $P$-$D$ transitions, within the HF approximation we expect them
to occur at the following energies:

\begin{widetext}
\begin{eqnarray}
\hbar\omega_{1}(3) & = & {\cal E}^{(e)}_{1D_1}-{\cal
  E}^{(e)}_{1P_1}+
2\,J^{(ee)}_{SP_1}-K^{(ee)}_{SP_1}-2\,J^{(ee)}_{SD_1}-K^{(ee)}_{SD_1} +\delta_{1}(3)
  \nonumber\\ \\
\hbar\omega_{2}(3) & = & {\cal E}^{(e)}_{1D_2}-{\cal
  E}^{(e)}_{1P_1}+
2\,J^{(ee)}_{SP_1}-K^{(ee)}_{SP_1}-2\,J^{(ee)}_{SD_2}-K^{(ee)}_{SD_2}+\delta_{2}(3) \nonumber\\ \\
\hbar\omega_{3}(3) & = & {\cal E}^{(e)}_{2S}-{\cal
  E}^{(e)}_{1P_1}+2\,J^{(ee)}_{SP_1}-K^{(ee)}_{SP_1}-2\,J^{(ee)}_{S2S}-K^{(ee)}_{S2S}
  + \delta_3(3)\nonumber\\
\end{eqnarray}
\end{widetext}

\noindent We find in our CI calculations [Fig. \ref{Fig_3}(a)] that the strong feature
around $49\;{\rm meV}$ corresponds to the (nearly overlapping) $1P$-$1D_1$ and
$1P$-$1D_2$ transitions. The weak transition at $\sim 52\;{\rm meV}$ arises
from $1P_1$-$2S$ and because $|2S\rangle$ in the alloyed dot is primarily
oriented (polarized) along $[110]$ the transition is weak.

%%%%%%%%%%%%%%%%%%%%%%%%
%   Intraband vs N_h   %
%%%%%%%%%%%%%%%%%%%%%%%%
%
%
\subsection{Valence transitions vs $N_h$}

Earlier calculations assumed simple models with the incorrect symmetry and
neglected the multiband nature of the hole single-particle states, which leads
to an incorrect treatment of the hole-hole interaction.
Within our atomistic approach, spin-orbit coupling and the multiband nature of
the hole single-particle states prevent us from writing meaningful HF
expressions in the case of {\em holes}. So we discuss directly the results of
our CI calculations. Figure \ref{Fig_3}(b) shows the valence intraband
transitions for $N_h=1$, $2$, and $3$ for light polarized along $\hat{\bf
  e}\parallel [1\bar 10]$. In general, compared to the conduction case, the
valence intraband spectra are more sensitive to the number of holes in the
dot.

{\em $N_h=2$:} The $1S$-$1P_1$ transition (lowest feature in $N_h=1$) appears
red-shifted by nearly $6\;{\rm meV}$ and split---two peaks between
$8$-$11\;{\rm meV}$.  Due to the hole-hole exchange interaction this
transition is split in a pair of low-energy, nearly doubly degenerate states
and two higher-lying states mutually split by $\sim 1\;{\rm meV}$. Similarly,
transition $1S$-$1P_2$ splits in two transitions: One transition at $\sim
14\;{\rm meV}$, with an ensuing red-shift of $6\;{\rm meV}$, and another at
$\sim 20\;{\rm meV}$ that appears slightly blue-shifted ($\sim 1\;{\rm meV}$)
with respect to the transition at $N_h=1$.

Note that contrary to the case of electrons, and due to the presence of
spin-orbit interaction, the four states arising from the two-hole
configuration $h^1_0h^1_2$ do not split in a triplet and one singlet. Instead,
these four states split in two doublets that are allowed under IR light
excitation. More importantly, in the commonly used EMA with two-dimensional
harmonic confinement and without spin-orbit coupling one would not find these
double-peak structure of allowed transitions, but instead one would find a
spectra that resembles that of the $N_e=2$ electron case.

{\em $N_h=3$:} While the ground state is well described by
$|h^2_0h^1_1\rangle$, the effect of configuration mixing in the final states
(upon absorption) due to hole-hole interaction is remarkably pronounced and
leads to a complex spectrum. As a result, it is not possible to determine
unambiguously the spectroscopic shifts $\Delta(3)$. Prominent features are the
following.

(i) The lowest-energy peak at nearly $9\;{\rm meV}$ corresponds to transition
$1S$-$1P_1$. Also, the peak at $\sim 15\;{\rm meV}$ is mainly $1S$-$1P_1$, but
mixed with $1P_1$-$2S$ and $1P_1$-$2S$. This mixing leads to the high
intensity of this transition. Remarkably, we find that in contrast to the
$N_e=3$ case, the $1S$-$1P_1$ transition {\em is not bleached} by having a
hole occupying the $1P_1$ state.

(ii) The peak at $10\;{\rm meV}$ correspond to transition $1S$-$1P_2$, while
the weaker feature at $\sim 11\;{\rm meV}$ is due to a transition with
$1P_1$-$1D_1$ predominant character.

(iii) Above $15\;{\rm meV}$ the features in the spectrum correspond to
transitions to heavily mixed final configurations: (a) The lower-energy peak
in the double-peak structure around $20\;{\rm meV}$ corresponds to a mixture
of the allowed $1S$-$1P_1$ and $1S$-$1P_2$ transitions in addition to a
sizeable component ($16\%$) of the forbidden $h^2_0h^1_1$-$h^1_0h^2_2$
transition. In turn, the higher-energy peak is a mixture of $1S$-$1P_2$ and
$1P_1$-$D_1$ transitions. (b) The peak at $26\;{\rm meV}$ arises from two
nearly overlapping transitions. These transitions are a mixture of allowed
$P$-$D$ and $P$-$F$ transitions, as well as forbidden transitions.

(iv) Although {\em significantly weaker} than the other features in the
spectrum, the peak at $24\;{\rm meV}$ corresponds to a forbidden transition
made allowed by configuration mixing with allowed transitions. We also have
found this type of transitions in the {\em interband} spectra of
(In,Ga)As/GaAs dots.\cite{narvaez_PRB_2006}

%%%%%%%%%%%%%%%
%             %
%   Summary   %
%             %
%%%%%%%%%%%%%%%
%
%
\section{Summary}

By combining an atomistic, pseudopotential-based approach with the
configuration method, we have calculated the conduction and intraband
transitions in (In,Ga)As/GaAs quantum dots with up to three carriers. 
We illustrated our calculations with a prototypical lens-shaped
In$_{0.6}$Ga$_{0.4}$As/GaAs dot with diameter $b=252\;${\AA} and height
$h=35\;${\AA}. And as a benchmark, for dots charged with a single carrier, we
provided results for a pure non-alloyed InAs/GaAs dot with the same size.
We have made specific predictions that could be probed in {\em single-dot}
infrared spectroscopy of {\em n}-doped and {\em p}-doped dot:

(i) In pure, non-alloyed InAs/GaAs dots, the $1S$-$1P$ conduction intraband
transitions are fully in-plane polarized, while valence transitions are weakly
  polarized because the hole {\em P} states do not show any in-plane
  preferential orientation.

(ii) In alloyed In$_{0.6}$Ga$_{0.4}$As/GaAs dots the in-plane polarization of
$1S$-$1P$ conduction intraband transitions strongly depend on alloy
fluctuations, which change the in-plane orientation of the nearly generate
P-shell states. The polarization of valence intraband transitions is
insensitive to changes in alloy fluctuations.

(iii) Upon changing the number of carriers in the dot, the intraband
transitions display spectroscopic shifts of about $1$-$2\;{\rm meV}$. These
shifts are not well described within Hartree-Fock, instead their magnitude is
determined by correlation effects.

(iv) Spin-orbit coupling and the multi-band characteristic of holes states result
in important differences between the {\em many-particle} valence and
conduction intraband spectra. Spectroscopic shifts can only be determined
unambiguously for conduction transitions.

\begin{acknowledgements}
This work was funded by the U.S. Department of Energy, Office of Science,
Basic Energy Sciences, under contract No. DE-AC36-99GO10337 to NREL, and by
NREL Director's DDRD program.
\end{acknowledgements}

\end{document}